\documentclass[fleqn,usenatbib]{mnras}

\pdfoutput=1
\usepackage{float}
\usepackage{subfiles}
\usepackage{graphicx}
\usepackage{enumitem}
\usepackage{amssymb}
\usepackage{amsmath}
\usepackage[lofdepth,lotdepth]{subfig}

\def\muhz{\mu\mathrm{Hz}}
\def\numax{\nu_{\mathrm{max}}}

\def\dnu{\Delta \nu}

\def\teff{T_{\mathrm{eff}}}

\def\feh{\rm{[Fe/H]}}

\def\teffsun{T_{\mathrm{eff,} \odot}}

\def\rsun{R_{\odot}}

\def\msun{M_{\odot}}
\def\numaxsun{\nu_{\mathrm{max,} \odot}}
\def\dnusun{\Delta\nu_{\odot}}

\def\Ks{K_{\mathrm{s}}}

\def\fdnu{f_{\Delta \nu}}
\def\fnumax{f_{\numax}}

\newcommand{\Kepler}{{Kepler}}

\newcommand{\Gaia}{{Gaia}}
\newcommand{\nablaeddy}{\nabla_{\mathrm{e}}}
\newcommand{\nablaad}{\nabla_{\mathrm{ad}}}
\newcommand{\nablarad}{\nabla_{\mathrm{rad}}}
\usepackage{hyperref}
\graphicspath{{plots/}{../plots/}}

\title[Adiabatic error in luminous giants]{The effect of the adiabatic assumption on asteroseismic scaling relations for luminous red giants}                                                                 
\author[J. C. Zinn et al.]{           
Joel C. Zinn,$^{1,2,3,4,5}$\thanks{E-mail: joel.zinn@csulb.edu} 
Marc H. Pinsonneault,$^{3}$
Lars Bildsten$^{4}$
and Dennis Stello$^{2,6,7,8}$
\\                                                                                                                         
$^{1}$Department of Astrophysics, American Museum of Natural History, Central Park West at 79th Street, New York, NY 10024, USA\\                           
$^{2}$School of Physics, University of New South Wales, Barker Street, Sydney, NSW 2052, Australia\\                                                 
$^{3}$Department of Astronomy, The Ohio State University, 140 West 18th Avenue, Columbus, OH 43210, USA\\
$^{4}$Kavli Institute for Theoretical Physics, University of California, Santa Barbara, CA 93106, USA\\
$^{5}$Department of Physics and Astronomy, California State University, Long Beach, Long Beach, CA 90840, USA\\
$^{6}$Sydney Institute for Astronomy (SIfA), School of Physics, University of Sydney, NSW 2006, Australia\\
$^{7}$Stellar Astrophysics Centre, Department of Physics and Astronomy, Aarhus University, Ny Munkegade 120, DK-8000 Aarhus C, Denmark\\
$^8$Center of Excellence for Astrophysics in Three Dimensions (ASTRO-3D), Australia
}                                                                                              
\date{Accepted XXX. Received YYY; in original form ZZZ}                                                    
\pubyear{2023}                                                                                            
\begin{document}                                                                                          
\label{firstpage}                                                                                                          
\pagerange{\pageref{firstpage}--\pageref{lastpage}}                                                                        
\maketitle        

\begin{abstract}
Although stellar radii from asteroseismic scaling relations agree at the percent level with independent estimates for main sequence and most first-ascent red giant branch stars, the scaling relations over-predict radii at the tens of percent level for the most luminous stars ($R \gtrsim 30 \rsun$). These evolved stars have significantly superadiabatic envelopes, and the extent of these regions increase with increasing radius. However, adiabaticity is assumed in the theoretical derivation of the scaling relations as well as in corrections to the large frequency separation. Here, we show that a part of the scaling relation radius inflation may arise from this assumption of adiabaticity. With a new reduction of Kepler asteroseismic data, we find that scaling relation radii and \Gaia\ radii agree to within at least $2\%$ for stars with $R \lesssim 30\rsun$, when treated under the adiabatic assumption. The accuracy of scaling relation radii for stars with $50\rsun \lesssim R \lesssim 100\rsun$, however, is not better than $10\%-15\%$ using adiabatic large frequency separation corrections. We find that up to one third of this disagreement for stars with $R \approx 100\rsun$ could be caused by the adiabatic assumption, and that this adiabatic error increases with radius to reach $10\%$ at the tip of the red giant branch. We demonstrate that, unlike the solar case, the superadiabatic gradient remains large very deep in luminous stars. A large fraction of the acoustic cavity is also in the optically thin atmosphere. The observed discrepancies may therefore reflect the simplified treatment of convection and atmospheres.
\end{abstract}


\section{Introduction}
\label{sec:intro}
\subsection{Adiabatic asteroseismic scaling relations for ensemble asteroseismology}
Ensemble asteroseismology has flourished following the space-based missions of CoRoT \citep{baglin+2006}, \Kepler\ \citep{borucki+2008}, and K2 \citep{howell+2014}, which have provided tens of thousands of asteroseismic detections of stars on the red giant branch (RGB) \citep[e.g.,][]{de-assis-peralta_samadi_michel2018,yu+2018,stello+2017}. With TESS \citep{ricker+2014} expected to deliver hundreds of thousands of RGB asteroseismic detections \citep{aguirre+2020,hon+2021}, and PLATO \citep{rauer+2014} also planned to deliver a similar number \citep{mosser+2019}, ensuring that asteroseismic parameters are accurate is crucial for stellar physics and Galactic archaeology.

The equations governing stellar pulsations are derived by linearizing stellar structure equations that have been perturbed around the equilibrium solution. This linearization has largely followed the adiabatic perturbative analysis of \cite{tassoul1980} (in the tradition of adiabatic approximations of \cite{pekeris1938} and \cite{cowling1941}). In this treatment, the fully general heat equation,
\begin{equation}
\label{eq:dqdt}
    \frac{dq}{dt} = \frac{1}{\rho (\Gamma_3 - 1)} \left ( \frac{dp}{dt} - \frac{\Gamma_1 p}{\rho}\frac{d\rho}{dt} \right ),
    \end{equation}
is traditionally approximated under the assumption that there is no heat gain or loss ($\frac{dq}{dt} = 0$), and so the adiabatic expression for how pressure and density vary is recovered:
\begin{equation}
    \frac{dp}{dt} = \frac{\Gamma_1 p}{\rho}\frac{d\rho}{dt}.
\end{equation}

The effect of the adiabatic approximation on inferred stellar parameters derived from asteroseismic inversions has been quantified for a $3.83\rsun$ RGB star by \cite{buldgen+2019}. An exploration for more luminous giants has not been done, nor has the adiabatic error been quantified for asteroseismic scaling relations, which are used in ensemble asteroseismic analysis of thousands of stars (e.g., from TESS or K2). Compared to asteroseismic inversions that use the frequencies of all observed modes to constrain stellar models, asteroseismic scaling relations yield stellar radius and mass using only two characteristic frequencies in a solar-like oscillator spectrum: $\dnu$ and $\numax$. 
    
Regarding $\dnu$, \cite{tassoul1980} showed that the solution of the resulting stellar pulsation equations have eigenvalue solutions (corresponding to mode frequencies) that are regularly separated in frequency space at large radial order.\footnote{ Although $\dnu$ may be defined for any degree, in what follows, we consider $\dnu$ as defined by radial modes: $\dnu \equiv \nu_{n,\ell=0} - \nu_{n-1,\ell=0}$.} In particular, the separation of modes of the same degree but adjacent radial order, is given by the so-called large frequency separation \citep{ulrich1986,kjeldsen&bedding1995}:

\begin{equation}
\frac{\dnu}{\dnusun} \approx \sqrt{\frac{M/\msun}{(R/\rsun)^3}}.
\label{eq:_scaling2}
\end{equation}

The frequency at maximum power, $\numax$, is known both empirically and theoretically to be related to the pressure scale height in the stellar atmosphere, and therefore to stellar surface gravity and effective temperature \citep{brown+1991,kjeldsen&bedding1995,chaplin+2008,belkacem+2011}:
\begin{equation}
\frac{\numax}{\numaxsun} \approx \frac{M/\msun}{(R/\rsun)^2\sqrt{(\teff/\teffsun)}}.
\label{eq:_scaling1}
\end{equation}

In what follows, we adopt $\numaxsun = 3076\muhz$ and $\dnusun = 135.146\muhz$ \citep{pinsonneault+2018}.

\subsection{Quantifying scaling relation errors with $\fdnu$ and $\fnumax$}
There has been a consensus in recent years that the observed $\dnu$ does not exactly follow the scaling relation \citep[e.g.,][]{stello+2009}, which has been understood to be because Equation~\ref{eq:_scaling2} is valid only when $\dnu$ corresponds to the frequency separation at infinitely large radial order. Of course, in practice, the observed modes in stars are at finite radial order. Though turbulent convective processes are responsible for exciting the modes, they also tend to damp high-frequency modes \citep{belkacem+2011}, meaning that the observed modes have radial order between 3-10 in RGB stars and between 10-30 in main sequence stars. This consideration has motivated a modified version of Equation~\ref{eq:_scaling2}:
\begin{equation}
\frac{\dnu}{\fdnu \dnusun} \approx \sqrt{\frac{M/\msun}{(R/\rsun)^3}},
\label{eq:scaling2}
\end{equation}
where $\fdnu$ captures differences in the observed $\dnu$ and the $\dnu$ that would be measured at infinite radial order. This is done by first matching a stellar structure model to an observed star, given a temperature, a metallicity, etc. A scaling relation $\dnu$ is computed directly from the model mean stellar density according to Equation~\ref{eq:_scaling2}. Model pulsation frequencies are then calculated from the stellar structure model using linearized, adiabatic stellar pulsation equations and a $\dnu$ is computed using the model frequencies at low radial order in the vicinity of the observed $\numax$ to create a model-observed $\dnu$. The model scaling relation $\dnu$ is then divided by the model-observed $\dnu$ to yield $\fdnu$.

Apart from the adiabatic approximation, this formalism assumes that $\fdnu$ will map the observed $\dnu$ onto what the observed $\dnu$ would be at high radial order, where the scaling relation between $\dnu$ and mean stellar density in Equation~\ref{eq:_scaling2} should hold. Following the first calculation of a temperature-dependent $\fdnu$ by \cite{white+2011}, other authors have identified metallicity \citep{guggenberger+2016}, and also mass- \& evolutionary state--dependences \citep{sharma+2016}. The corrections for solar-mass stars are typically around $1\%$ for main sequence stars and may reach $3\%$ for RGB stars with under the adiabatic approximation.

The $\numax$ scaling relation in Equation~\ref{eq:_scaling1}, unlike the one for $\dnu$ in Equation~\ref{eq:_scaling2}, is independent of the linearized stellar pulsation equations, depending instead on the local physics of the atmosphere (it still assumes adiabaticity via a dependence on the local adiabatic sound speed, however). The $\numax$ scaling relation is semi-empirical rather than exact, because predicting the spectrum of modes that will be observed requires a theoretical model of mode damping and excitation. This motivates another correction factor, $\fnumax$, that would encapsulate errors in the scaling relation itself due to non-adiabatic (or other) terms entering into Equation~\ref{eq:_scaling1} and/or systematics in the measurement of $\numax$:
\begin{equation}
\frac{\numax}{\fnumax \numaxsun} \approx \frac{M/\msun}{(R/\rsun)^2\sqrt{(\teff/\teffsun)}}.
\label{eq:scaling1}
\end{equation}

Unfortunately, $\numax$ cannot yet be computed from stellar models at the precisions required to calculate theoretical $\fnumax$ as is possible for $\fdnu$ (though see \citealt{zhou+2020} for promising progress). Here, we assume $\fnumax = 1$, though we discuss potential reasons for $\fnumax \neq 1$ in \S\ref{sec:fnumaxfdnu}.

Rearranging Equations~\ref{eq:scaling2}~\&~\ref{eq:scaling1} yields asteroseismic scaling relations for radii and masses:
\begin{align}
\frac{R}{\rsun} &\approx \left(\frac{\numax}{\fnumax \numaxsun}\right)
\left(\frac{\dnu}{\fdnu \dnusun}\right)^{-2}
\left(\frac{\teff}{\teffsun}\right)^{1/2},
\label{eq:radius}
\end{align}
and 
\begin{align}
\frac{M}{\msun} &\approx \left(\frac{\numax}{\fnumax \numaxsun}\right)^3
\left(\frac{\dnu}{\fdnu \dnusun}\right)^{-4}
\left(\frac{\teff}{\teffsun}\right)^{3/2}.
\label{eq:mass}
\end{align}

The correction to $\dnu$, $\fdnu$, thus enters into both the radius and mass scaling relation. The correction is especially important for asteroseismic ages, which, for low-mass RGB stars, would scale approximately like $\fdnu^{12}$, according to the mass-luminosity relationship for a solar-metallicity, solar-mass main sequence star \citep[e.g.,][]{salaris_cassisi2005}. 

Applying $\dnu$ corrections improves agreement between asteroseismic radii and independent radii from eclipsing binaries \citep{gaulme+2016a,brogaard+2018a} and \Gaia\ radii \citep{huber+2017} as well as between asteroseismic masses and astrophysical priors \citep{epstein+2014,sharma+2016}. A correction also made it possible to reconcile mass measurements in open clusters with scaling relations \citep{pinsonneault+2018}. However, the corrections do not seem to be as effective in the luminous giant regime $R \gtrsim 30\rsun$, where there are discrepancies between $10-20\%$ in radius compared to eclipsing binary radii \citep{kallinger+2018} and \Gaia\ radii \citep{zinn+2019rad}.

The most luminous stars have photometric variability large enough to be detected by ground-based surveys such as OGLE \citep[e.g.,][]{udalski+2008}, the All-Sky Automated Survey \citep{pojmanski1997}, the All-Sky Automated Survey for SuperNovae \citep{shappee+2014}, and the Zwicky Transient Facility \citep{bellm+2019}. With hundreds of thousands of light curves available for evolved RGB stars in ground-based observations, and hundreds of thousands more in space-based observations from TESS \citep[e.g.,][]{hon+2021}, an understanding of the observed pulsations will be enhanced by a proper theoretical treatment. Asteroseismology of the most luminous RGB stars --- luminous asteroseismology --- with these and future surveys like PLATO \citep{rauer+2014} and Rubin Observatory \citep{ivezic+2019} will allow Galactic archaeology studies out to tens of kiloparsecs \citep[e.g.,][]{auge+2020,hey+2023}, yielding distances to better precision than Gaia \citep[e.g.,][]{huber+2017} and asteroseismic ages for stars in the outer halo and outer disc.

Given the large problems in the radius scaling relation for luminous stars ($R \gtrsim 30\rsun$) noted in the literature, it is of great interest to ensure the accuracy of $\fdnu$ in the luminous giant regime. The adiabatic approximation, through its effect on $\fdnu$, is a potential area for improvement in this regard, and we explore the magnitude of its impact on $\fdnu$ in what follows.

\section{Methods}
\subsection{Stellar structure models}
\begin{figure*}
    \centering
    \includegraphics[width=0.65\textwidth,trim=0 80 0 100]{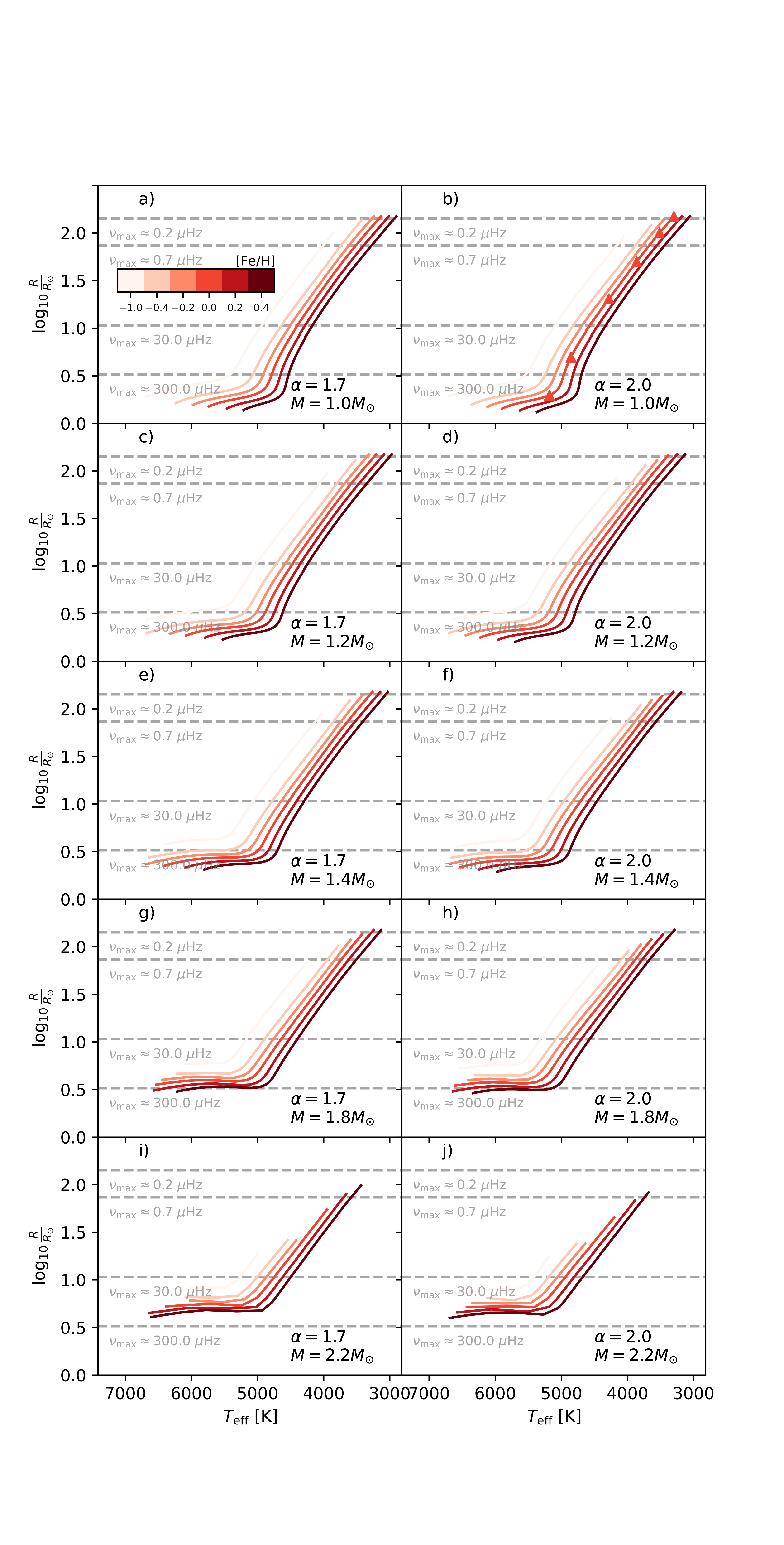}
    \caption{Stellar evolution tracks up to the RGB tip for all of the models considered here, where luminosity increases with increasing radius. The RGB bump is suppressed for plotting and interpolation purposes and models too hot to excite solar-like oscillations are also not shown. Each panel represents a different mass/mixing length parameter combination, while each track has a color corresponding to its metallicity according to the color bar. The triangles mark solar-mass, solar-metallicity, $\alpha=2.0$ models shown in Fig.~\ref{fig:deldiff}.}
    \label{fig:hr}
\end{figure*}

Asteroseismic frequency calculations require knowledge of the detailed stellar interior from stellar structure models. Our stellar structure models for the RGB stars considered here are run using version 15140 of MESA  \citep{paxton+2011a,paxton+2013a,paxton+2015a,paxton+2018a,paxton+2019}. Models were run without rotation, overshooting, diffusion, or mass loss. Convection was treated according to the \cite{cox_giuli1968} mixing length prescription. Opacities were generated from OPAL \citep{iglesias_rogers1993,iglesias_rogers1996} using \cite{grevesse+1998a} solar abundances, with C/O enriched abundance mixtures assumed for helium burning. In the low-temperature regime, molecular opacities are adopted from \cite{ferguson+2005}. An example parameter file used to compute our models is provided in Appendix A. We consider the evolution of RGB stars at two different mixing length parameters and a range of metallicities to gain a coarse understanding as a function of mass, metallicity, and mixing length of how the adiabatic assumption leads to errors in $\fdnu$ and therefore asteroseismic radius and mass. The Hertzsprung-Russell diagram of our models are shown in Figure~\ref{fig:hr}.

\subsection{GYRE non-adiabatic treatment}
We use GYRE \citep{townsend_teitler2013} to compute asteroseismic frequencies for our stellar models. GYRE may treat the stellar pulsation equations with or without the adiabatic approximation \citep{goldstein_townsend2020}, and does so by linearizing the energy transport equation, which we briefly review here.

We may write the left-hand side of Equation~\ref{eq:dqdt} as a balance of sources and sinks, i.e., a balance between energy generation per unit mass, $\epsilon$, and the energy flux, $\vec{F}$:
\begin{equation}
\label{eq:heat}
    \frac{dq}{dt} = \epsilon - \frac{1}{\rho} \nabla \cdot \vec{F}.
\end{equation}
In the above, $\vec{F}$ includes both convective and radiative fluxes.

GYRE linearizes the heat equation in Equation~\ref{eq:heat} and therefore may take into account non-adiabatic effects in the pulsation equations by allowing for pulsational perturbations of heat gain/loss, $\frac{dq}{dt}$. It does so according to:

\begin{equation}
    \frac{\partial \delta q}{\partial t} = \delta \epsilon - \frac{1}{\rho} \nabla \cdot \left( \vec{F_{\mathrm{rad}}}' + \delta r [\nabla \cdot \vec{F}_{\mathrm{rad}}]\right),
\end{equation}
where $\delta$ indicates a Lagrangian perturbation and a prime indicates an Eulerian perturbation. Here, $\delta r$ is the Lagrangian perturbation of the position and $\vec{F}_{\mathrm{rad}}$ is the radiative flux. The radiative flux is perturbed according to the diffusion approximation. The convective flux does not enter into the above linearization because GYRE operates under the frozen convection approximation, where the convective flux is not perturbed (meaning there is no coupling between convection and pulsation).

The parameter file used to compute stellar pulsation frequencies with GYRE is provided in Appendix B.

\subsection{Calculating adiabatic and non-adiabatic $\fdnu$}
The $\fdnu$ values are determined by first computing a model-observed $\dnu$ ($\dnu_{\rm{obs}}$), according to the procedure described in \cite{white+2011}, wherein the differences between radial frequencies are fitted with a least-squares approach and a weighting that is designed to mimic observational methods of determining $\dnu$. In detail, the weights are assigned according to a Gaussian centered on $\numax$, and with a $\numax$-dependent standard deviation, $\sigma$ taken from \citep{zinn+2019bam}: $\sigma = 1.05 \ln \numax - 1.91$. We compute $\dnu_{\rm{obs}}$ both under the adiabatic approximation ($\dnu_{\rm{obs,ad}}$) and in the non-adiabatic treatment ($\dnu_{\rm{obs,nonad}}$).

\begin{figure*}
    \centering
    \includegraphics[width=1.0\textwidth]{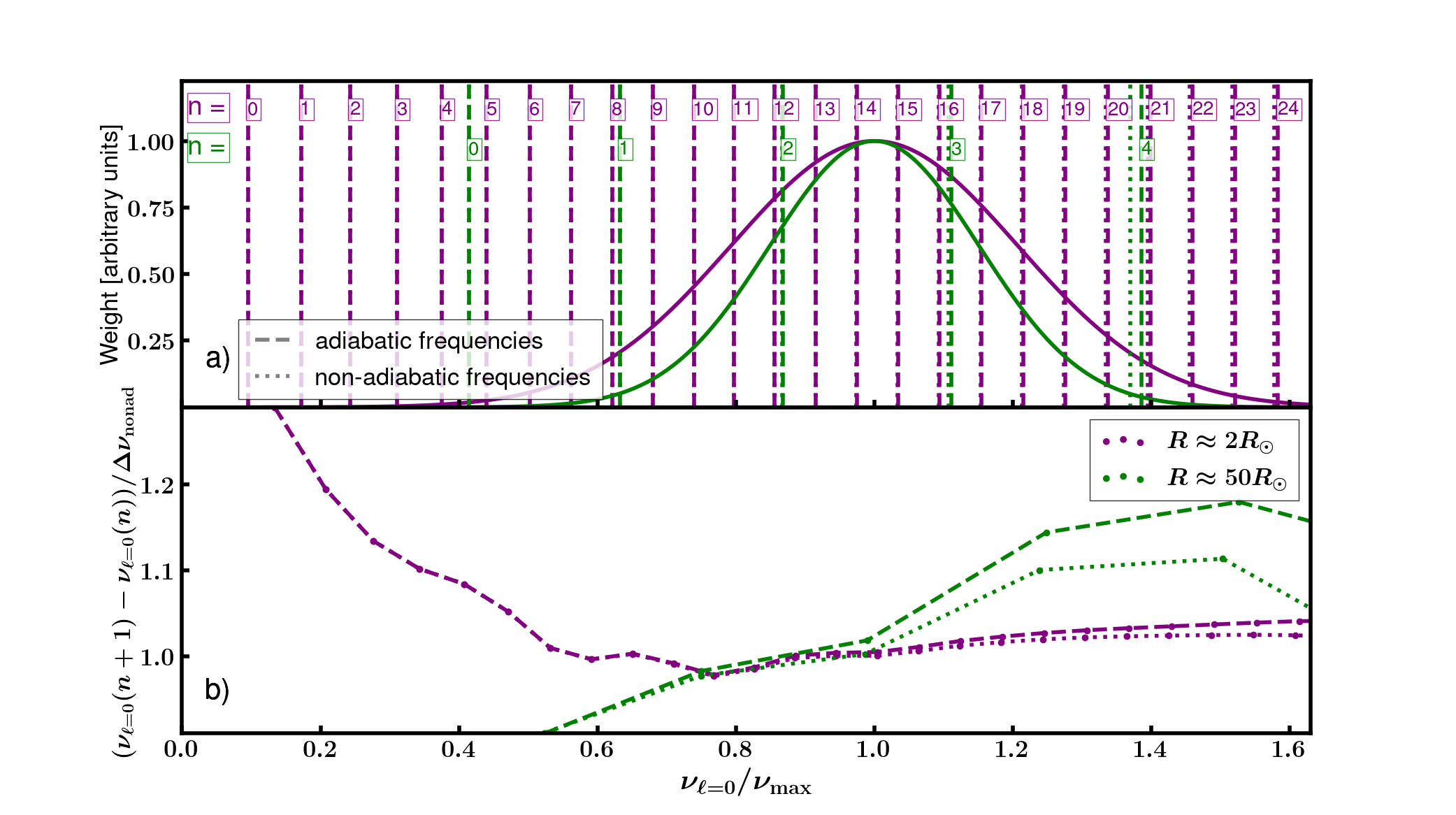}
    \caption{a) The modelled $\ell = 0$ frequencies for a solar-mass, solar-metallicity, $\alpha = 2.0$ star at $R \approx 2\rsun$ (purple vertical lines) and $R \approx 50\rsun$ (green vertical lines) computed with and without the adiabatic assumption (dashed and dotted lines, respectively), normalized by $\numax$. The radial order for each frequency is listed as $n$. The weighting scheme used to compute an average difference between successive modes is given by a Gaussian centered around $\numax$, the width of which depends on $\numax$ (solid curves). b) The difference between frequencies of adjacent radial order, $n$, as a function of frequency. The $R \approx 50\rsun$ star has many fewer modes around $\numax$ (number of vertical lines within the Gaussian in panel a) and $\dnu$ is affected much more by the adiabatic assumption (difference between green curves in panel b).}
    \label{fig:dnu}
\end{figure*}

We visually demonstrate the method for determining $\dnu$ in Figure~\ref{fig:dnu}. The weights for each of the frequency differences that go into $\dnu$ are indicated by the Gaussian in Fig.~\ref{fig:dnu}a. This Gaussian represents roughly the relative amplitude of the modes that would be measurable in surface brightness variation measurements. Note that the width of the Gaussian increases with $\numax$, encompassing more of the modes of the $R\approx 2\rsun$ model than of the $R \approx 50\rsun$ model, which only has a few modes that would be measurable. We also see in this panel that the difference between the adiabatic and non-adiabatic frequencies (dashed versus dotted vertical lines) increases with increasing frequency. This is due to the known behavior of higher-order radial modes to be more localized to the surface than lower-order radial modes: as we show in \S\ref{sec:physical}, the outer layers of evolved stars are the most superadiabatic, and so the adiabatic assumption is increasingly worse not only for evolved stars, but also for higher-order radial modes. We see the effect of this on the $\dnu$ measurement in Fig.~\ref{fig:dnu}b, which shows the difference between successive $\ell = 0$ frequencies. The final $\dnu_{\rm{obs,ad}}$ and $\dnu_{\rm{obs,nonad}}$ are the Gaussian-weighted averages of the respective curves shown here. The more evolved star shows significant departures between the adiabatic and non-adiabatic frequency differences with increasing frequency. Note that the y-axis here is normalized to the non-adiabatic $\dnu$, and so the effect is such that $\Delta \nu_{\mathrm{nonad}}$ is smaller than $\Delta \nu_{\mathrm{ad}}$.

The fractional error induced in the asteroseismic radius scaling relation (Eq.~\ref{eq:radius}) by calculating $\fdnu \equiv \dnu/\dnu_{\rm{obs}}$ in the adiabatic approximation is given solely by the ratio of $\dnu_{\rm{obs,ad}}^2/\dnu_{\rm{obs,nonad}}^2$ (see Equation~\ref{eq:radius}), which we abbreviate as $\dnu_{\rm{ad}}^2/\dnu_{\rm{nonad}}^2$ in what follows.

The effect of non-adiabaticity on $\dnu$ could plausibly be a function of bulk stellar properties (e.g., metallicity or mass). We therefore generalize this difference between $\Delta \nu_{\mathrm{nonad}}$ and $\Delta \nu_{\mathrm{ad}}$ across mass, metallicity, and mixing length, and as a function of radius in \S\ref{sec:results}.

\begin{figure*}
    \centering
    \includegraphics[width=0.65\textwidth,trim=0 80 0 100,clip]{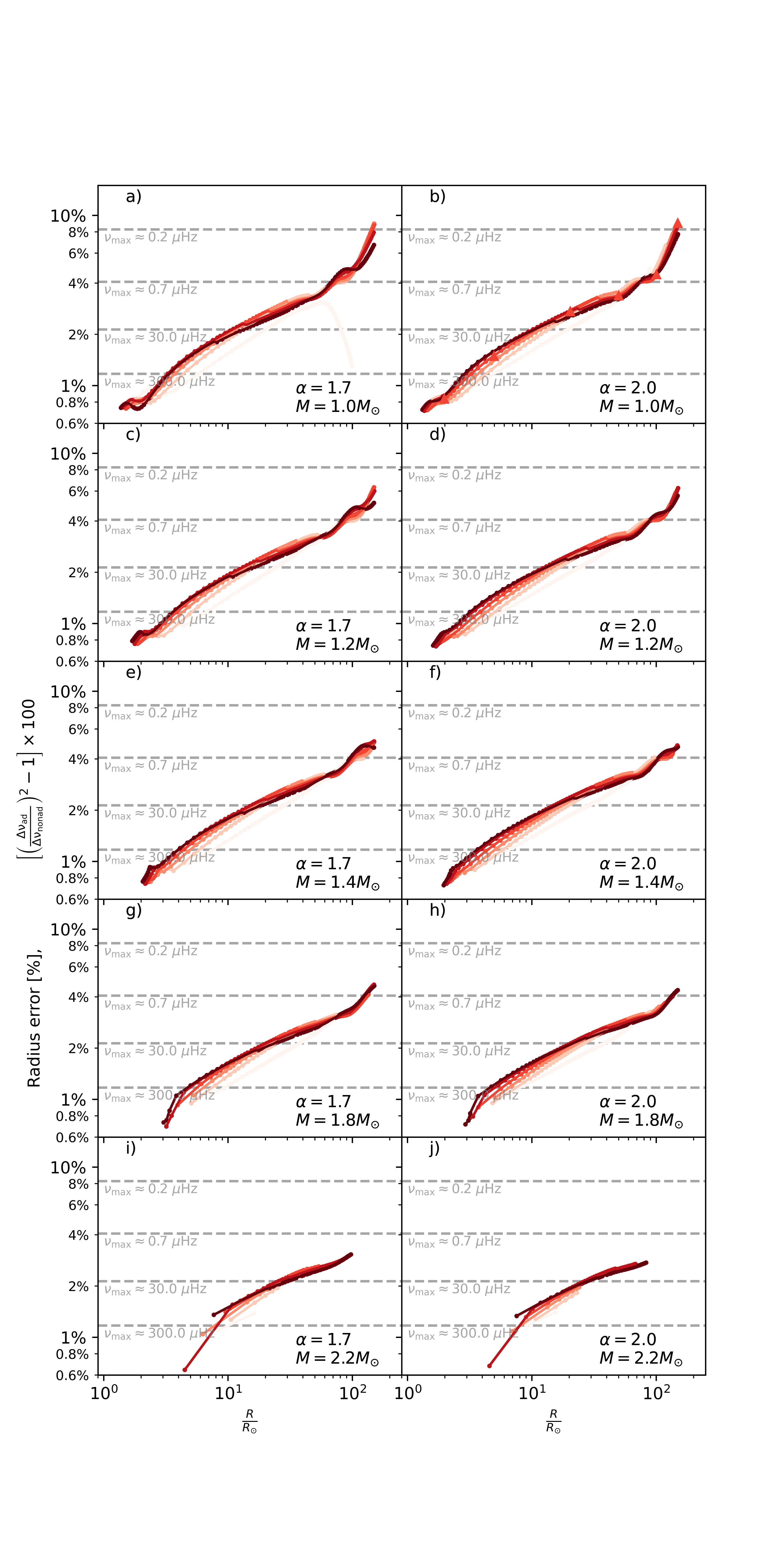}
    \caption{The expected error from using the asteroseismic radius scaling relation (Equation~\ref{eq:radius}) with theoretical corrections to observed $\fdnu$ in the adiabatic approximation (what we call the `adiabatic error'), in percent. Each point represents the error for stars of mass ($1\msun$, $1.2\msun$, $1.4\msun$, $1.8\msun$, $2.2\msun$), metallicities (-1, -0.4, -0.2, 0.0, 0.2, 0.4), and mixing length parameters (1.7, 2.0) at various stages of evolution along the RGB. The RGB bump is excluded, as are models with $R > 150\rsun$. Positive errors indicate that the asteroseismic radius is too large, which is consistent with the tension compared to independent \Gaia\ radii shown in Figure~\ref{fig:apokasc}. Each panel represents a different mass/mixing length parameter combination, while each track has a color corresponding to its metallicity according to the color bar of Fig.~\ref{fig:hr}. The triangles mark solar-mass, solar-metallicity, $\alpha=2.0$ models shown in Fig.~\ref{fig:deldiff}. Horizontal dashed lines give the approximate adiabiatic errors for the solar-mass, solar-metallicity, $\alpha = 2.0$ models at various $\numax$ values, for reference (in detail, $\numax$ depends on mass, metallicity, and $\alpha$). }
    \label{fig:err}
\end{figure*}

\section{Results and discussion}
\label{sec:results}

\subsection{The adiabatic error as a function of stellar mass, metallicity, and mixing length}
In Figure~\ref{fig:err} is shown the error that would be induced in the asteroseismic radii via a $\fdnu$ computed under the adiabatic approximation (called the adiabatic error in the following discussion). The error here is shown in percent, where a positive value would mean that a non-adiabatic $\fdnu$ reduces the asteroseismic radius. The effect is as large as $10\%$ near the tip of the RGB, and is less than $\approx 2\%$ for stars with $R \lesssim 30\rsun$.

Figure~\ref{fig:err} shows a modest metallicity-dependent error, and a few percent difference in the adiabatic error between the $M = 1\msun$ and $M = 1.8\msun$ models for models above $R \approx 100\rsun$; the $M = 2.2\msun$ model does not go as high up the RGB as the other models, but is consistent with the others up to $R \approx 100\rsun$. There also does not appear to be a significant difference between the error between the two mixing length parameters considered.

\begin{figure}
    \centering
    \includegraphics[width=0.5\textwidth]{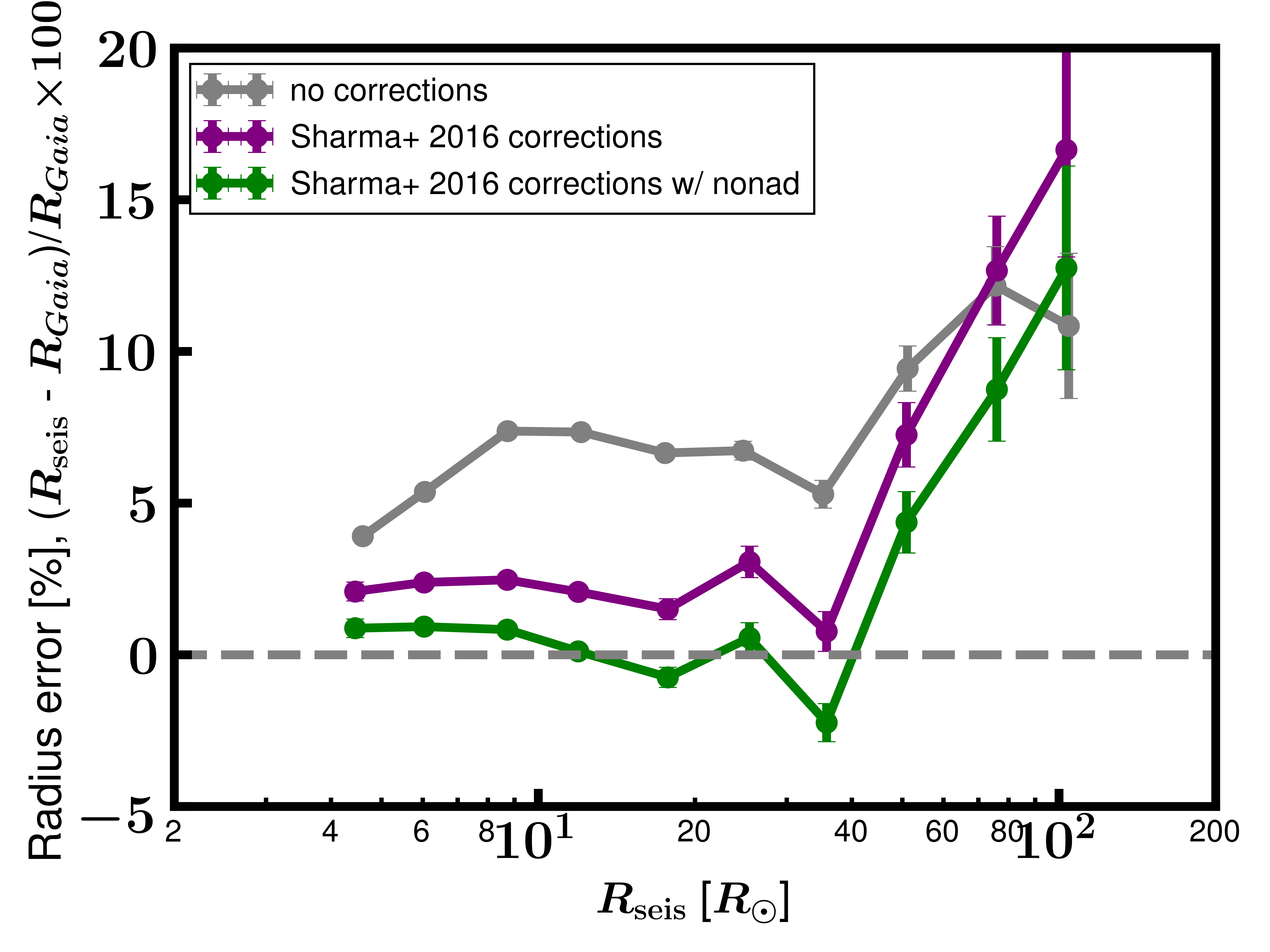}
    \caption{Fractional difference of asteroseismic and Gaia radii in the regime usually probed by asteroseismic studies ($R \lesssim 30\rsun$) and in the luminous giant regime ($R \gtrsim 30\rsun$) for stars with $ 1 < M/\msun < 2.2$ and $-2 < \feh < 0.4$. The binned medians and uncertainty on the medians of the fractional radius difference for raw asteroseismic scaling relation radii (Equation~\ref{eq:radius}, with $\fdnu = \fnumax = 1$) are shown in grey error bars. If the asteroseismic radii agreed with \Gaia, the data would follow the horizontal dashed line. The improvement from correcting $\dnu$ according to theoretical corrections from \protect\cite{sharma+2016} in the adiabatic regime are shown as purple error bars (Equation~\ref{eq:radius}, with $\fdnu \neq 1, \fnumax = 1$); though the agreement is improved at smaller radii, at radii $R \gtrsim 50\rsun$, the discrepancy persists and becomes worse than uncorrected asteroseismic radii. Correcting $\fdnu$ for non-adiabatic effects results in better agreement at large radii (green error bars). See text for details.}
    \label{fig:apokasc}
\end{figure}

\begin{figure}
    \centering
    \includegraphics[width=0.5\textwidth]{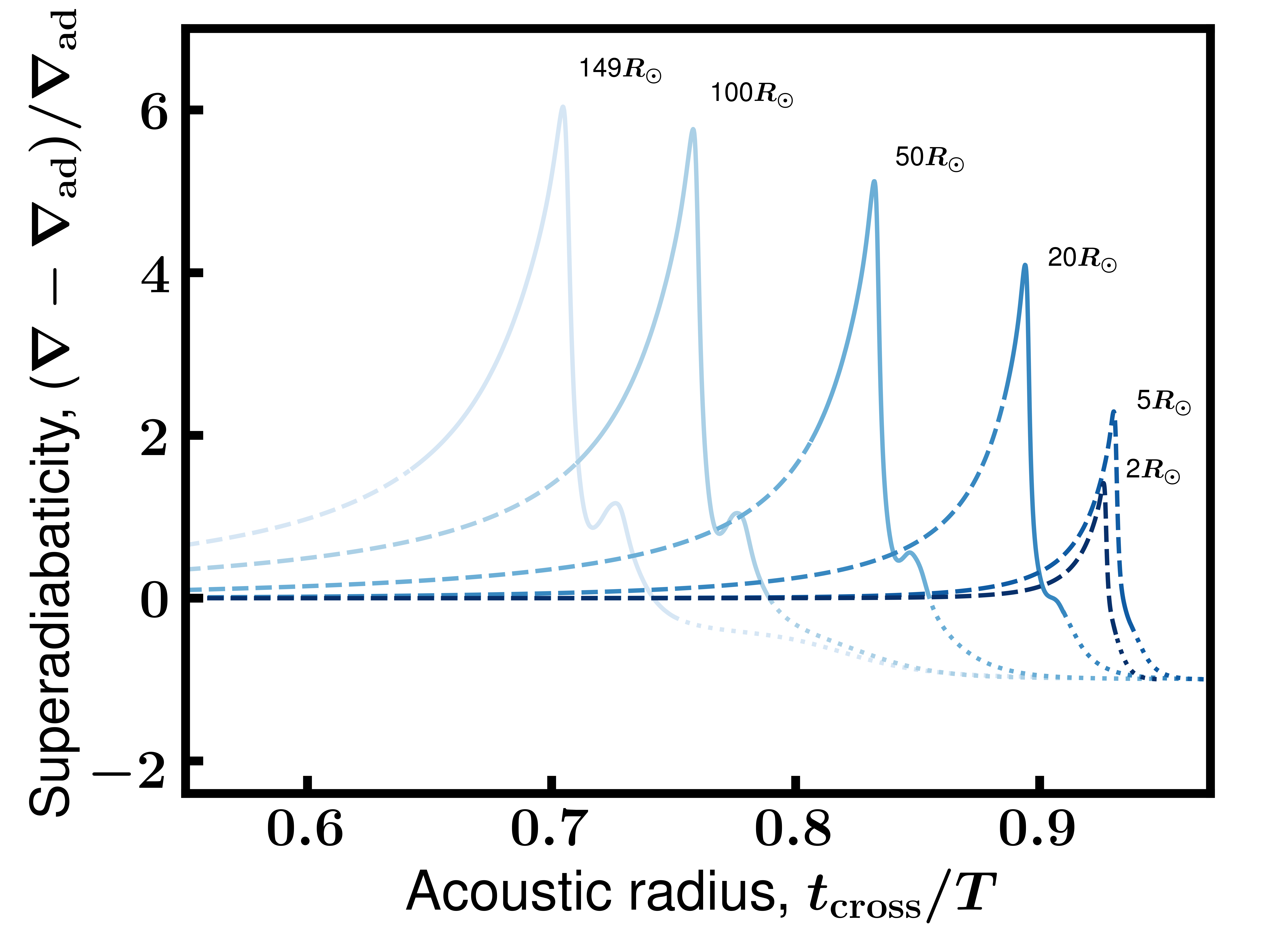}
    \caption{The fractional difference in the actual average temperature gradient, $\nabla$, from the adiabatic case, $\nablaad$, as a function of the acoustic radius (Equation~\ref{eq:acoustic}) for solar-mass, solar-metallicity, $\alpha = 2.0$ models at stellar radii of $\approx$  $5\rsun$, $2\rsun$, $20\rsun$, $50\rsun$, $100\rsun$, and $150\rsun$, from right to left. An acoustic radius of 0 corresponds to the stellar center, and one of unity corresponds to the stellar surface.  Dotted curves correspond to where MESA attaches an Eddington atmosphere. More evolved RGB stars have envelopes that are more strongly \text{superadiabatic} and are \text{superadiabatic} for a larger fraction of the acoustic cavity than those of less-evolved RGB stars, which may be partially responsible for a breakdown in the (adiabatic) asteroseismic scaling relations among RGB stars with $R \gtrsim 50\rsun$. In particular, that more of the superadiabatic region is in the regime where the thermal and sound crossing timescales have a ratio of $t_{\rm{th}}/t_{\rm{cross}} < 1$ (solid curves) motivates our relaxing the adiabatic assumption when computing theoretical asteroseismic frequencies. The acoustic radius of both the peak superadiabaticity and the top of the convection zone ($\nabla - \nablaad = 0$) move in step with each other, suggesting a relationship between the superadiabatic region expansion with increasing stellar radius and convection physics.}
    \label{fig:deldiff}
\end{figure}

\begin{figure}
    \centering
    \includegraphics[width=0.5\textwidth]{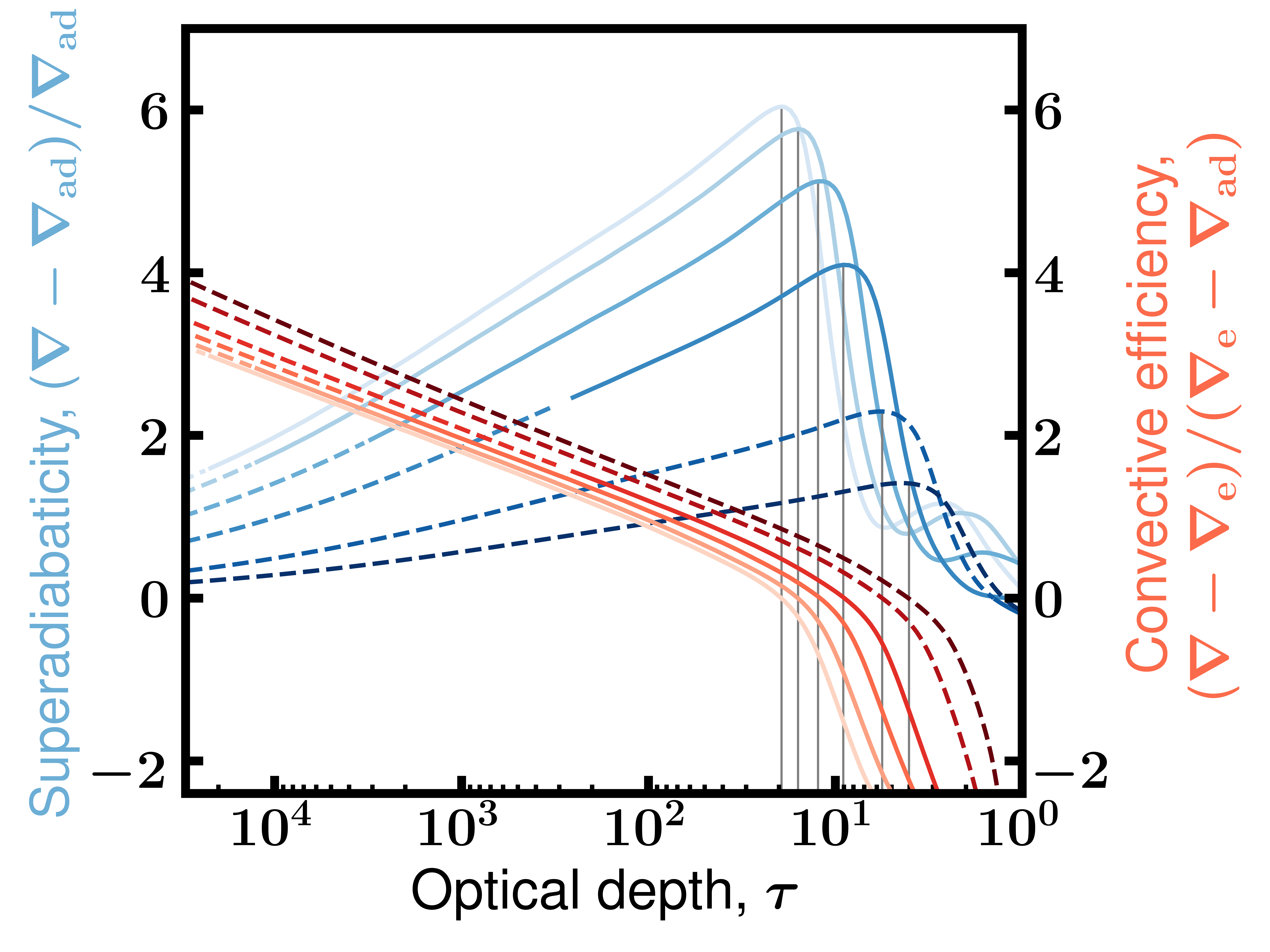}
    \caption{The fractional difference in the temperature gradient from the adiabatic case (left axis) and the efficiency of convection (right axis), as a function of optical depth for solar-mass, solar-metallicity, $\alpha = 2.0$ models at stellar radii of $\approx$ $2\rsun$, $5\rsun$, $20\rsun$, $30\rsun$, $100\rsun$, and $150\rsun$, from right to left. The red and blue curves of the same hue correspond to the same model. Convective efficiency less than 0 means that radiation dominates the energy transport, even if the region is still convective. Solid, dashed, and dotted parts of the superadiabaticity curves are the same as in Fig.~\ref{fig:deldiff}. Peaks in the superadiabaticity correspond well to where the convective efficiency equals 0 (solid vertical lines).}
    \label{fig:deldiffgamma}
\end{figure}

\begin{figure}
    \centering
    \includegraphics[width=0.5\textwidth]{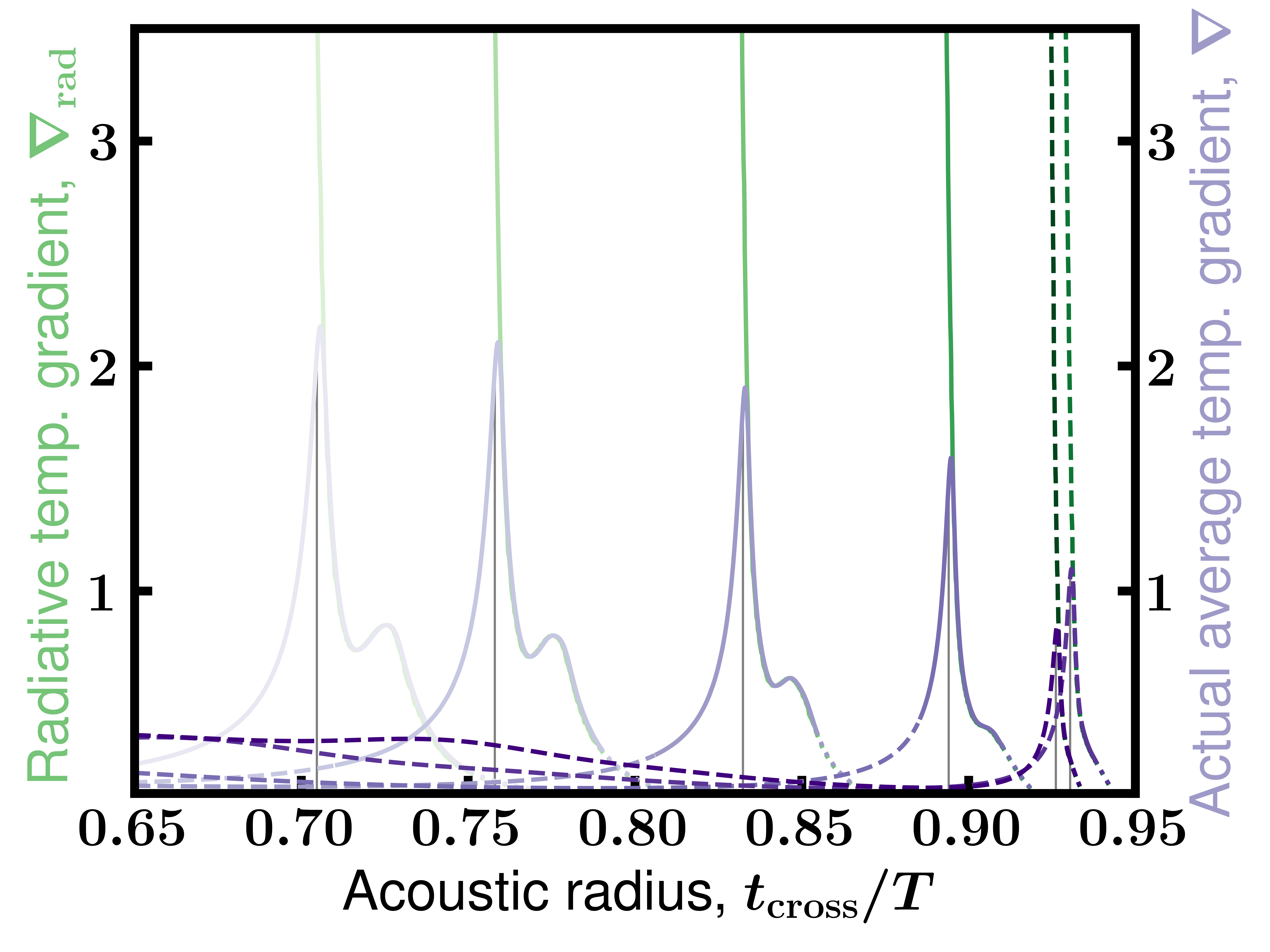}
    \caption{The radiative thermal gradient (left axis) and the actual average thermal gradient (right axis), as a function of acoustic radius for solar-mass, solar-metallicity, $\alpha = 2.0$ models at stellar radii of $\approx$ $5\rsun$, $2\rsun$, $20\rsun$, $30\rsun$, $100\rsun$, and $150\rsun$, from right to left. The purple and green curves of the same hue correspond to the same model. The peak of the superadiabaticity shown in the blue curves of Figs.~\ref{fig:deldiff}~\&~\ref{fig:deldiffgamma} is indicated by the vertical lines. Solid, dashed, and dotted parts of the superadiabaticity curves are the same as in Fig.~\ref{fig:deldiff}. The convective efficiency of the models decreases from small to large acoustic radius (from stellar interior to exterior), and, as it does, the actual average thermal gradient approaches the radiative thermal gradient, which describes the thermal gradient in the limit that all of the energy is transported via radiation. $\nabla = \nablarad$ corresponds to the maximum superadiabaticity in each model,  which then decreases, since $\nabla$ continues to track the radiative thermal gradient, which decreases with increasing acoustic radius. A small increase in $\nabla$ and $\nablarad$ exterior to the superadiabaticity maximum corresponds to the hydrogen partial ionization zone.}
    \label{fig:deldiffact}
\end{figure}

\subsection{Impact of the adiabatic error on the asteroseismic radius scale}
Having established that the adiabatic error will tend to inflate asteroseismic radii if not corrected, we now apply non-adiabatic corrections to the asteroseismic radius scale and compare the resulting radii with independent radii from \Gaia. The asteroseismic data we use for this are from APOKASC-3 (M. H. Pinsonneault et al., in preparation), with the \Gaia\ DR3 \citep{gaia-collaboration+2021} radii computed according to \cite{zinn+2017plxtgas} and \cite{zinn2021}, with \Gaia\ parallax corrections from \cite{lindegren+2021} and \cite{zinn2021}.\footnote{The comparison here is not sensitive to the \Gaia\ parallax zero-point, after corrections from \citep{lindegren+2021} are applied to the parallaxes. The \Gaia\ parallax zero-point uncertainty from  \cite{zinn2021} is negligible in its effect on the \Gaia\ radii and cannot explain the discrepancy, which has also been noted by \cite{kallinger+2018} for stars with $R \approx 30\rsun$.} The parallax statistical uncertainties are re-calibrated according to \cite{el-badry_rix_heintz2021}. The \Gaia\ radius calculation proceeds according to the Stefan-Boltzmann law, where a luminosity is computed from a $\Ks$ bolometric correction from \citep{gonzalezhernandez&bonifacio2009} and \Gaia\ parallax \citep{gaia-collaboration+2021,lindegren+2021b}, as well as extinctions from \cite{green+2019}, as implemented in \texttt{mwdust}\footnote{\url{https://github.com/jobovy/mwdust}} \citep{bovy+2016}. The uncertainties on the extinctions are assumed to be $0.08$mag, and effective temperatures are adopted from APOGEE DR16 \citep{ahumada+2020}. $\fdnu$ from \cite{sharma+2016} and \cite{asfgrid}\footnote{The \texttt{asfgrid} code is publicly available at \url{http://www.physics.usyd.edu.au/k2gap/Asfgrid/}} are computed using APOGEE effective temperatures and metallicities; evolutionary state classifications from M. H. Pinsonneault et al., in preparation; and asteroseismic surface gravity from Equation~\ref{eq:_scaling1}. The metallicities are corrected for non-solar abundances in the \cite{salaris+1993} approximation using APOGEE [$\alpha/M$].

We first show in Figure~\ref{fig:apokasc} the disagreement between \Gaia\ DR3 radii and asteroseismic radii computed using no $\dnu$ corrections ($\fdnu = 1$ in Eq.~\ref{eq:scaling1}; grey error bars). Asteroseismic radii computed with $\fdnu = 1$ are inflated compared to \Gaia\ radii at all $R > 4\rsun$, with a $\approx 10\%$ error near the tip of the RGB.

Looking to the purple error bars, we see that theoretical $\fdnu$ from \cite{sharma+2016} and \cite{asfgrid} calculated in the adiabatic approximation reduce the disagreement until $R \approx 50\rsun$ (purple error bars). This is the first indication that the agreement between adiabatic $\dnu$--corrected asteroseismic radii and independent \Gaia\ radii is very good up to a radius of $\approx 50\rsun$: the median agreement for stars with $10\rsun \leq R < 30\rsun $ and for $30\rsun \leq R < 50\rsun $ is within the systematic uncertainties of the asteroseismic radius scale ($\approx 2\%$; \citealt{zinn+2019rad}).\footnote{The absolute level of agreement in Figure~\ref{fig:apokasc} is uncertain at the $2\%$ level, and depends on the temperature scale, bolometric correction choices, and choices in the normalization of the scaling relations ($\numaxsun$ and $\dnusun$ in Eq.~\ref{eq:radius}; \citealt{zinn+2019rad}). That non-adiabatic corrections improve agreement in the luminous RGB regime --- where the disagreement between adiabatic asteroseismology and \Gaia\ is at the $10\%$ level --- holds true despite small shifts in these scales.} Previous indications from APOKASC-2 data \citep{pinsonneault+2018} suggested agreement for stars with $10\rsun \leq R < 30\rsun $ was $1.9\% \pm 0.6\%$ and for stars with $R > 30\rsun $ was $8.7\% \pm 0.9\%$ \citep{zinn+2019rad}, in the sense that asteroseismic radii were too large. This indicates a significant improvement in agreement with APOKASC-3 compared to APOKASC-2 for stars with $R > 30\rsun $. This improvement may be due to two changes implemented in APOKASC-3 compared to APOKSAC-2: 1) the APOKASC-3 analysis used data that were optimized for high-luminosity stars; and 2) the APOKASC-3 analysis used improved outlier rejection when combining $\numax$ and $\dnu$ results from different pipelines to yield the consensus $\numax$ and $\dnu$ values. Ultimately, whatever shortcomings there may be in the asteroseismic $\dnu$ scaling relation would therefore seem to affect the highest luminosity stars with $R \gtrsim 50\rsun$, leaving less evolved stars with scaling relation radii accurate to at least $2\%$. 

Nevertheless, the disagreement sharply increases to $10\%$ for $\dnu$-corrected radii for stars with $R \approx 50\rsun$, and the corrections actually aggravate the uncorrected radius error at larger radii to become $\approx 15\%$ for $R \approx 100\rsun$.

The disagreement we see therefore motivates considering if relaxing the adiabatic assumption in calculating mode frequencies can help. The green curve shows the radius agreement when we correct $\dnu$ by using $\fdnu$ from \cite{sharma+2016} multiplied by  $\dnu_{\rm{obs,nonad}}/\dnu_{\rm{obs,ad}}$ from our models, using linear interpolation as a function of metallicity, mass, and $\numax$. This correction takes into account the increase in $\dnu$ we find using non-adiabatic pulsation frequencies (e.g., Figs.~\ref{fig:dnu}~\&~\ref{fig:err}). We see that the radius agreement is indeed improved, though there still remain discrepancies at large radius. We discuss the physical reasons behind why the non-adiabaticity in luminous giants is important in \S\ref{sec:physical} and why relaxing the adiabatic assumption would be expected to improve the accuracy of asteroseismic radii. In \S\ref{sec:fnumaxfdnu}, we explore potential reasons for the remaining asteroseismic radius inflation even after correction for the adiabatic error.

It should also be noted that the inflated asteroseismic radius scale implies an inflated asteroseismic mass scale, as well: both Equation~\ref{eq:_scaling2}~\&~\ref{eq:_scaling1} depend on $\fdnu$ in the same sense. The mass dependence on $\fdnu$ is even stronger than that of radius, and so, fractionally, the inferred inflation of the asteroseismic mass scale is larger than the observed asteroseismic radius scale inflation seen in Figure~\ref{fig:apokasc}.

\subsection{Physical origins of non-adiabaticity in luminous giants}
\label{sec:physical}
The very good agreement in \Gaia\ and asteroseismic radius for the less luminous stars prompts us to consider the physical reasons behind the dramatic onset of asteroseismic radius inflation seen in Figure~\ref{fig:apokasc}.

We show in Figure~\ref{fig:deldiff} the fractional difference between the actual temperature gradient, $\nabla \equiv \frac{d\ln T}{d\ln P}$, and the temperature gradient in the adiabatic approximation, $\nabla_{\rm{ad}}$ as a function of the acoustic radius. Given the local sound crossing timescale, $t_{\rm{cross}}(r) \equiv \frac{dr}{c(r)}$, where $c(r)$ is the local adiabatic sound speed at radius, $r$, we define the acoustic radius to be 
\begin{align}
\label{eq:acoustic}
t_{\rm{cross}}/T &\equiv \int_r^\mathcal{R} \frac{dr}{c(r)}/\int_0^\mathcal{R}\frac{dr}{c(r)}\\
      &\equiv \int_0^\mathcal{R} dr \frac{\rho(r)} {\Gamma_{1}(r) P(r)}/T.
\end{align}
Here, $\Gamma_{1}$ is the local adiabatic exponent, $\rho$ is the local gas density, $P$ is the local pressure, and the integration limits are such that the acoustic radius increases from 0 at the stellar center to 1 at the stellar acoustic surface, $\mathcal{R}$. Note that the acoustic surface in general will differ from the photospheric surface \citep[e.g.,][]{balmforth_gough1990,lopes_gough2001,houdek_gough2007}. Here, we approximate the total sound crossing time as  $T = \frac{1}{2}\dnu^{-1}$ via the asymptotic expression for $\dnu$ \citep{tassoul1980}; the precise acoustic radius does not impact our discussion in what follows.

We can see in Figure~\ref{fig:deldiff} that the more evolved the first-ascent RGB star, the longer time a mode will travel through a \text{superadiabatic} region. This observation is only important, however, if the heat exchange occurs on a timescale that an asteroseismic mode experiences. The two relevant timescales are the local thermal timescale and the local sound crossing timescale. The local sound crossing timescale is given by Equation~\ref{eq:acoustic}. The local thermal timescale is the time taken to transport the energy of a shell of mass $dm = 4\pi r^2 dr \rho(r)$, given the local luminosity, $\ell(r)$. Given the internal energy per mass of a monatomic gas, $u = 3/2 \frac{kN_{\mathrm{A}}T}{\mu}$ and the ideal gas law $P = \frac{kN_{\mathrm{A}}\rho T}{\mu}$, we have that the local thermal timescale is
\begin{align}
t_{\rm{th}} &\equiv dm(r) u(r) / \ell(r)\\
    t_{\rm{th}} &= 4\pi r^2 dr \rho(r) 3/2 \frac{kN_{\mathrm{A}}T}{\mu} /\ell(r)\\
     &= 6 \pi r^2 dr P(r)/\ell(r),
\end{align}
so the ratio of the local thermal timescale to the local sound crossing timescale is
\begin{equation}
t_{\rm{th}}(r)/t_{\rm{cross}}(r) = 6 \pi r^2 P(r) c(r)/\ell(r).
\end{equation}

Solid curves in Figure~\ref{fig:deldiff} indicate regions where this ratio is below unity, corresponding to where the thermal timescale is shorter than the time taken for the mode to travel the region. This means that the mode will be sensitive to local heat exchange. To the extent that these regions correspond to superadiabatic regions in the star, the assumption of adiabaticity will be invalid for the linearized energy transport equation (Equation~\ref{eq:dqdt}). We see that the more evolved, larger RGB stars have a larger fraction of their acoustic radius in this regime (solid curves where the superadiabaticity is non-zero in Fig.~\ref{fig:deldiff}).

\begin{figure}
    \centering
    \includegraphics[width=0.5\textwidth]{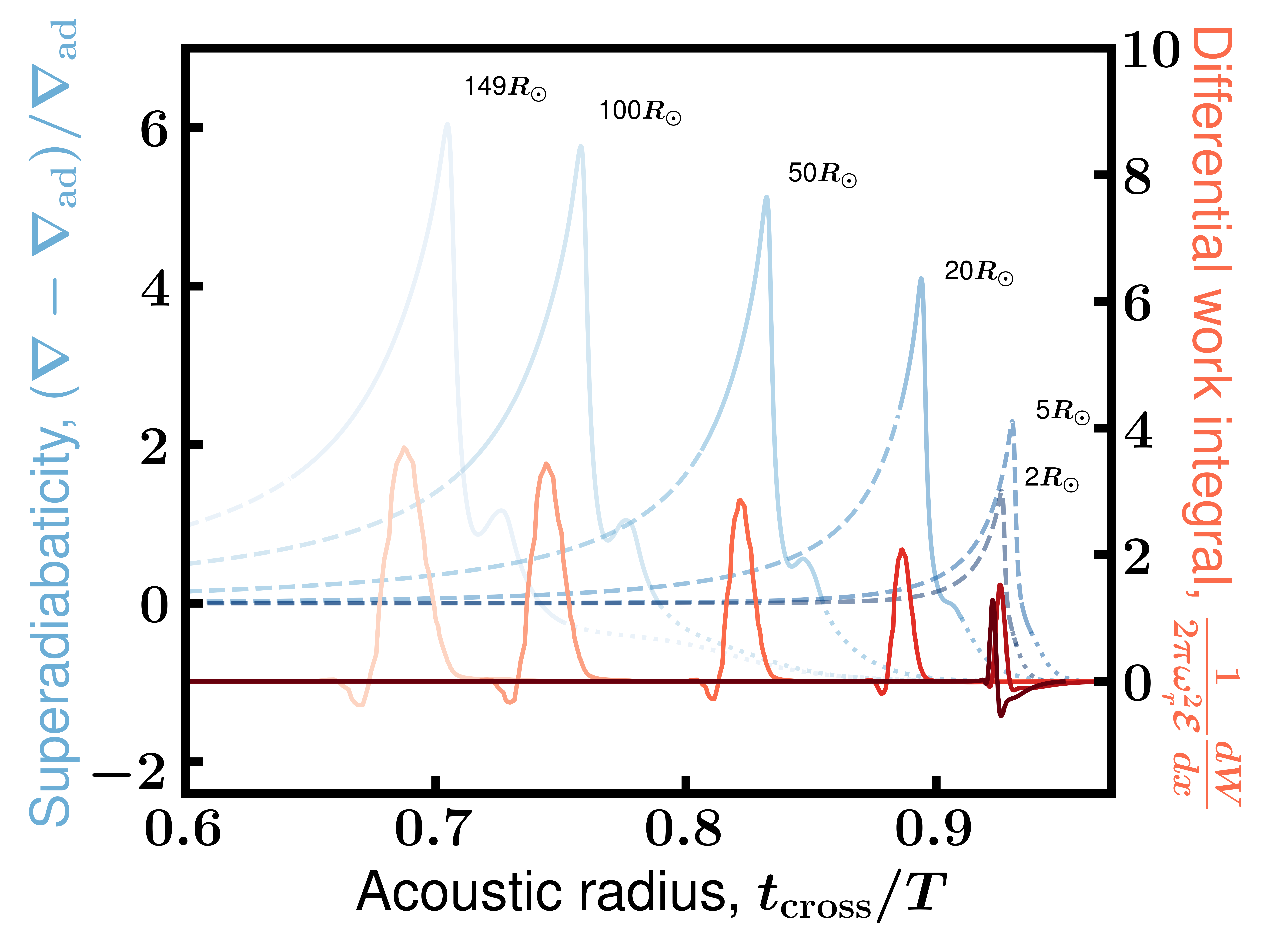}
    \caption{The superadiabaticity for solar-mass, solar-metallicity, $\alpha = 2.0$ models at stellar radii of $\approx$ $5\rsun$, $2\rsun$, $20\rsun$, $50\rsun$, $100\rsun$, and $150\rsun$, from right to left (as in Fig.~\ref{fig:deldiff}), compared to the corresponding differential work integrals for modes with radial order $n = 15, 10, 5, 4, 3, \&\ 2$, which correspond to frequencies near $\numax$. The normalization is chosen such that the integral of the normalized differential work integral from the center of the star ($x = 0$) to the surface ($x = 1$) yields the ratio of the imaginary to real components of the mode's eigenfrequency, and depends on the real component of the mode's eigenfrequency, $\omega_r$ (i.e., the observed, dimensionless mode frequency), and the mode's inertia, $\mathcal{E}$.}
    \label{fig:work}
\end{figure}

The timescale considerations here are useful to identify the location and extent of regions where modes will be affected by non-adiabatic effects. We discuss at the end of this section the work integral \citep[e.g.,][]{aerts_christensen-dalsgaard_kurtz2010}, which describes the relative amounts of energy gain/loss that a mode may experience in these regions.

Looking at Figure~\ref{fig:deldiff}, we see that for less evolved stars (smaller radii), the superadiabaticity peaks at larger acoustic radii; this peak occurs at smaller acoustic radii for more evolved stars (larger radii). At the same time, the top of the convection zone (where $\nabla - \nablaad = 0$) similarly moves deeper into the star with increasing radius. This connection between a convective property of the models (the extent of the convection zone) and the location of peak superadiabaticity motivates investigating the convection properties of the RGB stars as causing superadiabaticity. 

We start by considering the convective efficiency, following \cite{cox_giuli1968}:

\begin{equation}
\label{eq:efficiency}
    \Gamma \equiv \frac{\nabla - \nabla_{\mathrm{e}}}{\nablaeddy - \nablaad},
\end{equation}
where $\nablaad$ is the adiabatic thermal gradient, $\nabla$ is the actual average thermal gradient, and $\nablaeddy$ is the thermal gradient of a convective element. The above assumes the convective element produces no energy, and only loses it to its environment through radiation. In the limit of efficient convection (large $\Gamma$), the fraction of energy carried by convection will approach unity, as long as the radiative thermal gradient is sufficiently larger than the adiabatic thermal gradient. In the limit of inefficient convection ($\Gamma \leq 0$), the fraction of energy carried by convection goes to zero as the actual average thermal gradient and the convective element thermal gradient both approach the radiative thermal gradient. 

The convective efficiency behavior shown in red hues in Figure~\ref{fig:deldiffgamma} explains qualitatively the behavior of the superadiabaicity shown in blue hues in Figure~\ref{fig:deldiffgamma}. This figure shows that the convective efficiency decreases with decreasing optical depth, which reaches zero at an optical depth that also corresponds to a peak in the superadiabaticity. Note that our optical depth, $\tau$, is defined by the Rosseland mean opacity, $\kappa$: $\tau(r) = \int_r^R\rho \kappa dr$. We see in Figure~\ref{fig:deldiffact} that the optical depth of maximum superadiabaticity occurs when convection is inefficient enough that heat is mostly transported via radiative diffusion, and thus the actual thermal gradient (purple curves) approaches the radiative thermal gradient (green curves). This optical depth is related to the critical optical depth $\tau_{\mathrm{crit}} = (P_{\mathrm{rad}}/P_{\mathrm{gas}})c/v_{\mathrm{conv}}$ described in \cite{goldberg_jiang_bildsten2022}, where $v_{\mathrm{conv}}$ is the characteristic velocity at which convective elements travel; $c$ is the speed of light; and $P_{\mathrm{rad}}$ \& $P_{\mathrm{gas}}$ are the radiation and gas pressures. This critical optical depth can be shown to correspond to when the convective efficiency (Equation~\ref{eq:efficiency}) $\Gamma \approx 1$, and therefore tracks the optical depth of the peak in the superadiabaticity we see in our models, which occurs for $\Gamma \approx 0$ (vertical lines in Fig.~\ref{fig:deldiffgamma}).  

The amplitude and extent of the superadiabatic region differ between less and more evolved stars because $\tau_{\mathrm{crit}}$ is smaller for less evolved stars (Fig.~\ref{fig:deldiffgamma}), which also corresponds to a smaller radiative thermal gradient. This means less evolved stars have a smaller $\nablarad$ that the actual thermal gradient needs to reach, and therefore a smaller amplitude of the peak superadiabaticity. Since the slope of the convective efficiency with optical depth is the same no matter how evolved the stars is (Fig.~\ref{fig:deldiffgamma}), the physical extent of the superadiabatic region is also smaller for less evolved stars, since the actual thermal gradient needs to climb --- at the same rate as a more evolved star --- to a smaller value.

In Fig.~\ref{fig:work}, we show the superadiabaticity overlaid with a differential work integral, $\frac{dW}{dx}$, normalized such that, when integrated from the center ($x = 0$) to surface ($x = 1$), it would equal the ratio of the imaginary to real components of the mode eigenfrequencies for modes near the scaling relation $\numax$ for the star (see Eq. 25.19 of \citealt{unno+1989} for the normalization factor). This normalization ensures that the curves quantify the non-adiabaticity of the modes since a larger magnitude of the imaginary frequency compared to the real frequency indicates a more non-adiabatic mode. The structure of the differential work integral closely traces the observed superadiabatic gradient that we have explored earlier in this section: the amplitude, width, and location of the superadiabaticity peaks correspond well to the amplitude, width, and location of the extrema in the differential work integral. This supports our interpretation of the superadiabaticity as a useful proxy for the qualitative effect of non-adiabaticity on the mode frequencies.

We note that GYRE predicts the modes to be driven (i.e., $\int_0^1 \frac{dW}{dx} dx > 0$) and not damped as expected for solar-like oscillations. The detailed balance between pulsational energy gain and loss is known to be sensitive to the treatment of convection \citep[e.g.,][]{baker_gough1979}. Although it is not possible in the current implementation of GYRE to include a time-dependent perturbation to the convective flux, it is, however, possible to explore an additional time-dependent term to the radiative flux (assumed by default in GYRE to follow the time-independent diffusion approximation) via the Eddington approximation \citep{unno_spiegel1966}. This approximation correctly yields the behavior of the radiative flux in the limit of low optical thickness, while preserving the correct behavior of the diffusion approximation at high optical thickness. \cite{guenther1994} shows for the case of the Sun that the diffusion approximation (adopted by default in GYRE) makes different predictions for the work integral and mode eigenfrequencies compared to those when adopting the Eddington approximation. We find that altering GYRE's treatment of the perturbation of the radiative flux from the default diffusion approximation to include a time-dependent term via the Eddington approximation does not appreciably change our results. Further analysis of the thermodynamics --- in particular the effects of a time-dependent convective flux --- would be a fruitful pursuit for continued work on non-adiabatic pulsations in evolved giants.

\subsection{Luminous giant asteroseismic scaling relation errors: $\fnumax$ or $\fdnu$?}
\label{sec:fnumaxfdnu}
The adiabatic error in $\fdnu$ of $\approx 4\%$ for stars with $R \approx 100\rsun$ we find (Fig.~\ref{fig:err}) still leaves a $\approx 5\%-10\%$ margin of unexplained systematic difference in asteroseismic radii for stars with $R \gtrsim 50\rsun$ compared to \Gaia\ radii (green curve in Fig.~\ref{fig:apokasc}). The adiabatic error we have demonstrated is thermodynamic in nature, but other contributions to an error in $\fdnu$ could very well arise due to structural problems in the model.

For instance, as we have shown, large parts of the acoustic cavity exhibit low convective efficiency (Fig~\ref{fig:deldiffgamma}), and would therefore require a different treatment of convection than mixing length theory can provide. Existing work has shown that 1D mixing length theory yields meaningfully different stellar structure in the outer layers of a star compared to outer layer structures from 3D treatments of convection \citep[e.g.,][]{trampedach+2017,jorgensen+2019,mosumgaard+2020}. The resulting effect on asteroseismic frequencies is in addition to errors resulting from neglecting the coupling of convective motions to asteroseismic modes, which requires an in-depth treatment of the turbulent and convective physics in the convective envelope \citep[e.g.,][]{balmforth1992,xiong_cheng_deng1997,grigahcene+2005}. Structural surface effects have been considered for the RGB case \citep[e.g.,][]{trampedach+2017} as have effects due to time-dependent convection \citep[e.g.,][]{xiong2021}, but both effects have not yet been considered together. To date, the combination of these effects on asteroseismic frequencies, to our knowledge, has only been studied in the solar case, which yields excellent agreement with observed frequencies \citep{houdek+2017}.

Another structural issue that may contribute to errors in $\fdnu$ is the assumption of a 1D, plane-parallel, Eddington atmosphere, which becomes increasingly inaccurate to describe the $\teff-\tau$ relation in the atmosphere of the star. Indeed, we see that the atmosphere (dotted curves in Fig.~\ref{fig:deldiff}) occupies more and more of a star's acoustic radius with increasing stellar radius. The result is that the atmospheric temperature and opacity structure of the atmosphere would increasingly impact the mode structure and frequencies for more evolved stars.

There has been an effort to correct for these surface effects empirically, via so-called surface corrections. Surface corrections are used in grid-based asteroseismic modelling wherein observed individual frequencies are fit using models to derive stellar parameters \citep[see, e.g.,][and references therein]{serenelli+2017a}, which is in contrast to the scaling relations of interest to us here.  Surface corrections take the form of power laws as a function of frequency, the parameters of which are fit on a star-by-star basis to bring modelled and observed frequencies into the best possible agreement. Surface corrections of $\sim 1 \%$ are required to bring modelled and observed frequencies into alignment among main sequence stars \citep{kjeldsen_bedding_christensen-dalsgaard2008} and up to $2\%$ in subgiants \citep{ball_gizon2017}. Supporting findings from \cite{sonoi+2015}, \cite{li+2018} find typical surface corrections of order $0.3\%$ among RGB stars more evolved than $R \sim 10\rsun$. \cite{li+2022} similarly suggest the impact of the surface correction on $\dnu$ is negligible in the luminous giant regime. Recent work has demonstrated a computation of $\fdnu$ using the grid-based surface correction formalism \citep{li+2022,li+2023}, which results in a $\approx 4\%$ downward shift in radii for red giant stars with $R \lesssim 20\rsun$, in the same sense as the $\approx 2\%$ shift we observe in Fig.~\ref{fig:apokasc}. A similar exploration of a surface correction $\fdnu$ is warranted in the luminous regime, given that the frequency shifts due to the superadiabatic gradient that we find here should behave as a surface term corrigible by a surface correction, and which presumably accounts for a part of the $4\%$ radius correction \cite{li+2023} find. Indeed, the grid-based surface correction formalism would also be expected to account for other sources of error in the physics of the surface layers apart from the adiabatic assumption (e.g., thermal structure due to mixing length theory and atmospheric physics mentioned in this section).

Furthermore, there are several reasons to believe that systematics in $\numax$ are responsible for a non-negligible part of the luminous giant asteroseismic radius problem. First, it is plausible that there are measurement systematics in $\numax$ that are separate from any theoretical problems in the scaling relations themselves. This is because, for luminous giants, there are only a handful of modes that are visible (e.g., $R \approx 50\rsun$ case in Fig.~\ref{fig:dnu}), which invalidates the approximation most asteroseismic pipelines make to measure $\numax$ (viz., the modes can be modelled with a Gaussian in frequency space). There is also a possibility that the granulation among these high-luminosity RGB stars behaves differently than on the lower RGB. Choices in the model for the contribution of granulation to observed power spectra of solar-like oscillators can affect $\numax$ measurements at the percent level in lower RGB stars \citep{kallinger+2014}, and this may be aggravated at low frequency due to potential changes in the underlying `true' model of granulation in these luminous stars and/or due to changes in how the models fit in the regime where there are relatively few data points to constrain the granulation parameters.

Apart from measurement issues implicated in luminous giant $\numax$, there could be problems in the $\numax$ scaling relation itself that would cause a $\fnumax \neq 1$. \cite{viani+2017}, for instance, propose a metallicity-dependent term to the $\numax$ scaling relation, based on the logic that the $\numax$ scaling relation is a statement about the pressure scale height and the sound speed at the surface of a star, which depends on the mean molecular weight. More detailed theoretical motivations for $\fnumax$ that are analogous to theoretically-computed $\fdnu$ will need to await improved modelling of the excitation and damping effects in models \citep[e.g.,][]{zhou+2020}.

\subsection{Comparison to other work}
\cite{buldgen+2019} computed surface correction effects due to the adiabaticity assumption for a low-luminosity first-ascent RGB star with radius $3.83\rsun$. Their analysis differs from ours in the crucial aspect that theirs was an inversion exercise, using individual mode frequencies to derive a stellar density, which was then compared to the truth. Their Fig. 10 shows that the adiabatic frequency approximation is a negligible error contribution, when considering n=1-20 modes. Ultimately, our results are not comparable to theirs, given we are interested in errors introduced in scaling relations, not when using inversion techniques.
 
The onset of the thermal timescale becoming shorter than the sound crossing timescale (solid curves in Fig.~\ref{fig:deldiffgamma}) is also consistent with a 3D simulation of the stellar envelope of a $2\msun$, solar metallicity RGB star with $\log g = 1$ from \cite{ludwig_kucinskas2012}. Our 1D MESA model of a $1\msun$ RGB star with $50\rsun$ model has the same surface gravity as their model. As shown in Figure~\ref{fig:deldiffgamma}, its thermal--sound crossing timescale transition occurs at $\tau \sim 3000$ (transition from solid to dashed in the fourth curve from the right), and in the \cite{ludwig_kucinskas2012} 3D model this transition occurs at $\tau \sim 1000$.  \cite{ludwig_kucinskas2012} conclude that the structure from their 3D simulation cannot be described using a stellar structure from 1D mixing length theory, which further motivates investigating the impact of structural errors in the convective envelope on $\fdnu$.

We also note that 3D simulations of red supergiants by \cite{goldberg_jiang_bildsten2022} indicate critical optical depths at which convective efficiency $\Gamma \approx 1$ of $\tau_{\mathrm{crit}} \approx 200-300$. Figure~\ref{fig:deldiffgamma} shows that the convective efficiency reaches unity in our tip of the RGB model at $\tau_{\mathrm{crit}} \approx 200$, as well (leftmost red curve in Figure~\ref{fig:deldiffgamma}). This similarity in $\tau_{\mathrm{crit}}$ is owing to similar ratios of radiation to gas pressures ($\approx 3\times 10^{-3}$) and similar convection velocities ($\approx 3 \mathrm{km}\,\mathrm{s}^{-1}$).

Regarding the small metallicity dependence of the adiabatic error in Figure~\ref{fig:err}, \cite{epstein+2014} found that scaling relation masses, corrected for $\fdnu$, of metal-poor ($\feh < 1$) stars are too massive by $11\% \pm 4\%$. Using a metallicity-dependent $\dnu$ correction, \cite{sharma+2016} found metal-poor stars are too massive by only $4\% \pm 5\%$. This would imply that metal-poor asteroseismic radii are over-estimated by of order $6 \%$ ($2-3\%$ according to \citealt{sharma+2016}).  We do not find evidence of this for our $\feh = -1$ models compared to other metallicities. In fact, the $\feh = -1$ model $\fdnu$ are consistently less affected by the adiabatic assumption than other metallicities, no matter the mass or mixing length parameter (lightest curves in Fig.~\ref{fig:err}).

\section{Concluding remarks}
\label{sec:conc}
Current asteroseismic scaling relations break down on the upper RGB. We find that a large fraction of the acoustic radius (up to 20\%) is superadiabatic in luminous stars, which has a significant impact on the model-dependent corrections to the $\dnu$ component of the asteroseismic scaling relations. The adiabatic error is found to be no larger than a couple percent for stars with radii below $\approx 30 \rsun$. This is consistent with the percent-level bounds in asteroseismic radius accuracy set here and by previous work using independent constraints from \Gaia\ \citep{zinn+2019rad}. However, the adiabatic error we predict reaches 10\% at the tip of the RGB. Empirical constraints on the asteroseismic radius scale from fundamental \Gaia\ radii (Fig.~\ref{fig:apokasc}) indicate that asteroseismic radii have errors in excess of this level for stars with $R \gtrsim 50\rsun$, which suggests additional sources of errors in the asteroseismic scaling relations. 

Although non-adiabatic effects improve the agreement between asteroseismic and fundamental radii, there are still clear discrepancies that remain. We have demonstrated that luminous giants are strongly superadiabatic deep into the envelope, and that a large fraction of the acoustic cavity is in the atmosphere. As a result, predicted model frequencies will be sensitive to the treatment of convection theory and the adopted surface boundary conditions; a grid-based surface correction approach as described in \cite{li+2023} may be well equipped to correct for both these and the non-adiabatic surface effects of the sort we have quantified here.  It is also likely that the $\numax$ component of the scaling relations is responsible for systematic errors in luminous giant asteroseismic radii. Unfortunately, theoretical predictions for $\numax$ that might permit a $\fnumax$ correction to observed $\numax$ (in analogy with $\fdnu$), are not yet precise enough \citep{zhou+2020} to quantify the $\numax$ contribution to asteroseismic radius errors.

Nevertheless, APOKASC-3 data indicate that stars with $R \lesssim 30\rsun$ have radii that agree to \Gaia\ within at least $2\%$ (Fig.~\ref{fig:apokasc}), which would allow for accurate asteroseismology of stars more luminous than typically analyzed in the literature.

This work has considered radius errors, but errors in $\dnu$ and $\numax$ will also impact asteroseismic masses, which are crucial data in Galactic archaeology and stellar physics studies \citep[e.g.,][]{miglio+2013,rendle+2019,sharma+2019,miglio+2021}. One may reduce the impact of unknown, outstanding errors in the mass asteroseismic scaling relations due to $\numax$ by using external radius measurements from \Gaia\ in combination with non-adiabatic $\dnu$ to yield hybrid \Gaia-asteroseismic masses.

\bibliographystyle{mnras}  
\bibliography{bib}

\section*{Acknowledgments}
We thank Richard Townsend for guidance in GYRE's non-adiabatic capabilities and Jared Goldberg \& Evan Anders for helpful discussions. We also thank the referees for their work in improving the manuscript. JCZ was supported by an NSF Astronomy and Astrophysics Postdoctoral Fellowship under award AST-2001869. JCZ and MHP acknowledge support from NASA grants 80NSSC18K0391 and NNX17AJ40G. DS is supported by the Australian Research Council
(DP190100666). This research was supported by NSF ACI-1663688 and PHY1748958, and by NASA ATP-80NSSC18K0560 and ATP-80NSSC22K0725.

Funding for the Stellar Astrophysics Centre (SAC) is provided by The Danish National Research Foundation (Grant agreement no. DNRF106). 

This publication makes use of data products from the Two Micron All Sky Survey, which is a joint project of the University of Massachusetts and the Infrared Processing and Analysis Center/California Institute of Technology, funded by the National Aeronautics and Space Administration and the National Science Foundation.

This work has made use of data from the European Space Agency (ESA)
mission
{\it Gaia} (\url{https://www.cosmos.esa.int/gaia}), processed by the
{\it Gaia}
Data Processing and Analysis Consortium (DPAC,
\url{https://www.cosmos.esa.int/web/gaia/dpac/consortium}). Funding
for the DPAC
has been provided by national institutions, in particular the
institutions
participating in the {\it Gaia} Multilateral Agreement.

Funding for the Sloan Digital Sky Survey IV has been provided by the Alfred P. Sloan Foundation, the U.S. Department of Energy Office of Science, and the Participating Institutions. SDSS-IV acknowledges
support and resources from the Center for High-Performance Computing at
the University of Utah. The SDSS web site is www.sdss.org.

\section*{Data Availability} 
Data used in this work are available upon reasonable request to the corresponding author.

\onecolumn

\appendix
\section{MESA inlist}
\begin{verbatim}
&kap
use_Type2_opacities = .true.
Zbase = 4.3d-2
/

&eos
/

&star_job
show_log_description_at_start = .false.
create_pre_main_sequence_model = .true.
save_model_when_terminate = .true.
save_model_filename = 'final.mod'
write_profile_when_terminate = .true.
filename_for_profile_when_terminate = 'final_profile.data'
/

&controls
use_dedt_form_of_energy_eqn = .true.	
use_gold_tolerances = .true.
mesh_delta_coeff = 0.5
use_other_mesh_functions = .true.
x_ctrl(1) = 500    
x_ctrl(2) = 0.02d0 
x_ctrl(3) = 0d0 
max_years_for_timestep = 1d6           
varcontrol_target = 1d-3
max_timestep_factor = 2d0
delta_lgT_cntr_limit = 0.1  
delta_lgRho_cntr_limit = 0.5
num_trace_history_values = 2	
trace_history_value_name(1) = 'rel_E_err'
trace_history_value_name(2) = 'log_rel_run_E_err'
photosphere_r_upper_limit = 1.5d2
mixing_length_alpha = 1.7
initial_mass = 1.2
initial_z = 4.3d-2
write_pulse_data_with_profile = .true.
pulse_data_format = 'GYRE'
format_for_FGONG_data = '(1p,5(E16.9))'	   
add_center_point_to_pulse_data = .false.
atm_option = 'T_tau'
atm_T_tau_relation = 'Eddington'
atm_T_tau_opacity = 'varying'	
initial_y = 0.28
cool_wind_RGB_scheme = ''
cool_wind_AGB_scheme = ''
RGB_to_AGB_wind_switch = 1d-4
Reimers_scaling_factor = 0.7d0  
Blocker_scaling_factor = 0.7d0  
cool_wind_full_on_T = 1d10
hot_wind_full_on_T = 1.1d10
hot_wind_scheme = ''
/

&pgstar
/
\end{verbatim}

\section{GYRE inlist}
\begin{verbatim}
&model
model_type = 'EVOL'
file = 'profile.data.GYRE'
file_format = 'MESA'
/

&constants                                                                                                                  
G_GRAVITY = 6.6740800000e-08                                                                                             
R_sun = 6.958e10                                                                                                         
M_sun = 1.988435e33                                                                                                      
/                                                                                                                                
&mode
l = 0
tag = 'radial'
/

&osc
outer_bound = 'VACUUM'
nonadiabatic = .TRUE.
tag_list = 'radial'
alpha_thm = 1	
/

&rot
/

&num
diff_scheme = 'MAGNUS_GL2'
nad_search = 'MINMOD'                                                                                           
restrict_roots = .FALSE.
/

&scan
grid_type = 'LINEAR'
freq_min_units = 'NONE'                                                                                                 
freq_max_units = 'NONE'
freq_min = 1.0 
freq_max = 40.0
n_freq = 2000  
tag_list = 'radial'
/

&grid
w_osc = 10
w_exp = 2 
w_ctr = 10
/

&shoot_grid
/

&recon_grid
/

&ad_output
summary_file = 'profile.data.GYRE.gyre_ad.eigval.h5'
summary_file_format = 'HDF'                            
summary_item_list = 'M_star,R_star,L_star,l,n_pg,n_g,omega,freq,E,E_norm,E_p,E_g' 
detail_template = 'profile.data.GYRE.gyre_ad.mode-%J.h5'           
detail_file_format = 'HDF'                   		   
detail_item_list = 'M_star,R_star,L_star,m,rho,p,n,l,n_p,n_g,omega,freq,E,E_norm,W,x,V,As,U,c_1,Gamma_1,nabla_ad,delta,xi_r,xi_h,phip,dphip_dx,delS,delL,delp,delrho,delT,dE_dx,dW_dx,T,E_p,E_g'
freq_units = 'UHZ'
/

&nad_output
summary_file = 'profile.data.GYRE.gyre_nad.eigval.h5'
summary_file_format = 'HDF'                            
summary_item_list = 'M_star,R_star,L_star,l,n_pg,n_g,omega,freq,E,E_norm,E_p,E_g' 
detail_template = 'profile.data.GYRE.gyre_nad.mode-%J.h5'  
detail_file_format = 'HDF'                   		   
detail_item_list = 'M_star,R_star,L_star,m,rho,p,n,l,n_p,n_g,omega,freq,E,E_norm,W,x,V,As,U,c_1,Gamma_1,nabla_ad,delta,xi_r,xi_h,phip,dphip_dx,delS,delL,delp,delrho,delT,dE_dx,dW_dx,T,E_p,E_g'
freq_units = 'UHZ'
/
\end{verbatim}

\bsp    
\label{lastpage}                      
\end{document}